\documentclass[runningheads,a4paper]{llncs}

\usepackage{times}
\usepackage{amssymb}
\usepackage{graphicx}
\usepackage{url}

\usepackage[T1]{fontenc}
\usepackage{amsmath}
\usepackage{makecell}

\newcommand{\keywords}[1]{\par\addvspace\baselineskip
\noindent\keywordname\enspace\ignorespaces#1}

\urldef{\mailsa}\path|{dominik.wagner,ilja.baumann,tobias.bocklet}@th-nuernberg.de|
\begin{document}
\title{Vocoder-Free Non-Parallel Conversion of Whispered Speech With Masked Cycle-Consistent Generative Adversarial Networks}
\titlerunning{Vocoder-Free Non-Parallel Conversion of Whispered Speech}
% If the paper title is too long for the running head, you can set
% an abbreviated paper title here
%
% \author{First Author\inst{1}\orcidID{0000-1111-2222-3333} \and
% Second Author\inst{2,3}\orcidID{1111-2222-3333-4444} \and
% Third Author\inst{3}\orcidID{2222--3333-4444-5555}}
% \author{Dominik Wagner\inst{1}\orcidID{0009-0006-7293-9131} \and Ilja Baumann\inst{1}\orcidID{0000-0002-1991-3710}  \and Tobias Bocklet\inst{1}\orcidID{0009-0008-7780-8821} 
% }
\author{Dominik Wagner \and Ilja Baumann \and Tobias Bocklet}

\authorrunning{D. Wagner et al.}
% First names are abbreviated in the running head.
% If there are more than two authors, 'et al.' is used.
%
\institute{Technische Hochschule Nürnberg Georg Simon Ohm, 90489 Nürnberg, Germany \\ \mailsa
% \email{\{dominik.wagner,ilja.baumann,tobias.bocklet\}@th-nuernberg.de}
}
\index{Wagner, Dominik}
\index{Baumann, Ilja}
\index{Bocklet, Tobias}

\toctitle{} \tocauthor{}

\maketitle

\begin{abstract}
Cycle-consistent generative adversarial networks have been widely used in non-parallel voice conversion (VC). 
Their ability to learn mappings between source and target features without relying on parallel training data eliminates the need for temporal alignments. 
However, most methods decouple the conversion of acoustic features from synthesizing the audio signal by using separate models for conversion and waveform synthesis. 
This work unifies conversion and synthesis into a single model, thereby eliminating the need for a separate vocoder. 
By leveraging cycle-consistent training and a self-supervised auxiliary training task, our model is able to efficiently generate converted high-quality raw audio waveforms. 
Subjective listening tests showed that our unified approach achieved improvements of up to 6.7\% relative to the baseline in whispered VC.
Mean opinion score predictions also yielded stable results in conventional VC (between 0.5\% and 2.4\% relative improvement). 

\end{abstract}
\section{Introduction}
%%\vspace{-2mm}
Voice conversion (VC) is the task of transforming an individual's voice into another, while preserving the linguistic and prosodic content \cite{sisman21vcoverview}. 
The conversion of whispered speech into voiced speech can be interpreted as a special case of VC, where the task is the recovery of the fundamental frequency ($F0$), without changing the linguistic content of an utterance, i.e., the whispered input is mapped to a corresponding normally phonated output produced by the same speaker. 
Many VC models, rely on the availability of parallel speech data, i.e., utterances with the same linguistic content are required from both source and target. 
However, the collection of such data is costly and often impractical. 

Furthermore, most approaches operating on parallel data require external time alignment, which may be imperfect and can heavily influence the conversion results \cite{helander08alignment,mohammadi17vcoverview}. 
Consequently, much effort has been placed on removing the parallel data requirement from VC systems. 
VC methods on non-parallel training data include autoencoders (AEs) \cite{qian19autovc,chou19adainvc,qian20aevc,lin21fragmentvc}, generative adversarial networks (GANs) \cite{kaneko18cycleganvc,kaneko19cycleganvc2,kaneko20cycleganvc3,kaneko21maskcyclevc,bac22nvcnet}, and more recently diffusion models \cite{popov2022diffvc,liu2021diffsvc}. 
Most VC systems reduce the resolution of the raw audio waveform to a lower-dimensional acoustic feature representation. 
In such cases, an additional vocoder model is necessary to reconstruct the waveform. 
These restrictions impede the development of efficient and specialized systems for low-resource tasks such as the conversion of voices from laryngeal cancer patients. 
To overcome these restrictions, we focus on GANs for non-parallel VC, that neither rely on parallel utterances and time alignment procedures nor on a secondary model for waveform synthesis. 
The proposed method is designed to produce high-quality waveforms from whispered speech features but can also be applied to conventional VC.\footnote{Audio samples and source code are available at:\\ \scriptsize{\url{https://audiodemo.github.io/voice-conversion}}} 

GAN-based non-parallel voice conversion shares similarities with image-to-image translation \cite{isola16patchgan,zhu17cyclegan,choi18stargan}, which is the task of finding mappings between images in a source domain and images in a target domain without the need for parallel training data \cite{sisman21vcoverview}.
%The goal is to we would like to transform one voice to that of another, while preserving the linguistic, and prosodic content. 
% A prominent model solving...
A prominent example for a successful attempt to solve the image-to-image translation task is CycleGAN \cite{zhu17cyclegan}. 
The success of cycle-consistent training on images has led to several adaptations, which aim to find optimal mappings between non-parallel speech data \cite{kaneko18cycleganvc,fang18cyclegan,kaneko19cycleganvc2,kaneko20cycleganvc3,kaneko21maskcyclevc}.
One of the first adaptations was CycleGAN-VC \cite{kaneko18cycleganvc}, which employs three loss functions: adversarial loss, cycle-consistency loss, and identity-mapping loss, to learn forward and backward mappings between source and target speakers. 
An improved version of CycleGAN-VC \cite{kaneko19cycleganvc2} incorporates a second adversarial loss, a generator that uses both 1D and 2D convolutional layers, and an improved discriminator based on PatchGAN \cite{li16patchgan,isola16patchgan}. 
In \cite{kaneko21maskcyclevc}, an auxiliary task consisting of randomly masking frames of the
input mel-spectrogram is added to the training procedure, to further improve the results. 
An adversarial network capable of directly performing voice conversion on the raw audio waveform was recently proposed in \cite{bac22nvcnet}. 
However, the speaker encoder component of their NVC-Net still relies on mel-spectrograms as input features. 

The advantages of cycle-consistent training have also been leveraged for whispered speech conversion \cite{parmar19whisper,patel20cincwhisper,malaviya20mspecnet}. 
However, these approaches model cepstral and $F0$ features with separate GANs and require additional vocoders for speech synthesis. 
Other methods for whispered speech conversion use the attention mechanism \cite{gao23attguided} to learn time alignments between utterance pairs and train models on data that was aligned via dynamic time warping (DTW) prior to training \cite{meenakshi18whsp,wagner22whisper,Wagner2024}. 
Some studies avoid time alignment by generating artificial whispers from voiced input speech \cite{pascual18whisper} or by representing the speech content in latent space before decoding it back to a mel-spectrogram representation \cite{rekimoto23wesper}. 
%%\vspace{-3mm}
\section{Method}
%%\vspace{-2mm}
Our method is inspired by MaskCycleGAN-VC \cite{kaneko21maskcyclevc} and its predecessors \cite{kaneko18cycleganvc,kaneko19cycleganvc2,kaneko20cycleganvc3}, as well as HiFi-GAN \cite{kong20hifigan}. 
We first introduce those systems before we describe our method in more detail. 
%%\vspace{-2mm}
\subsection{HiFi-GAN}
%%\vspace{-1mm}
HiFi-GAN \cite{kong20hifigan} is a waveform synthesis method consisting of a generator and multiple discriminator blocks. 
Multi-period discriminators (MPDs) are used to provide adversarial feedback based on disjoint audio samples, and multi-scale discriminators (MSDs) provide feedback based on the waveform at different resolutions. 
Apart from the standard adversarial loss terms \cite{goodfellow14gan} for generator and discriminator, HiFi-GAN employs an additional mel-spectrogram loss, as well as a feature matching loss in the generator objective. 
The generator uses transposed convolutions to upsample mel-spectrogram features until the length of the output sequence matches the temporal resolution of the raw waveform. 
A multi-receptive field fusion (MRF) component consisting of multiple residual blocks \cite{he16resnet}, with varying kernel sizes and dilations is used to generate features based on different receptive fields in parallel. 
The output of the MRF module is the sum  of the outputs of all residual blocks. 
%\vspace{-2mm}
\subsection{MaskCycleGAN-VC}
%\vspace{-1mm}
The architecture of MaskCycleGAN-VC \cite{kaneko21maskcyclevc} is based on CycleGAN-VC2 \cite{kaneko19cycleganvc2}. 
The generator combines 1D and 2D convolutional layers, as well as instance normalization \cite{ulyanov16instancenorm}, residual blocks \cite{he16resnet}, and gated linear units (GLUs) \cite{dauphin17glu}, to implement downsampling and upsampling of the input features. 
Input features are downsampled and passed through 6 residual blocks, before they are upsampled again. 
Downsampling is achieved via 2 blocks consisting of 2D convolutional layers, followed by instance normalization, and GLU activation. 
Each residual block consists of 1D convolutional layers, instance normalization and GLU activation. 
The 2 upsampling blocks employ 2D convolutional layers, instance normalization, GLU activation, and sub-pixel convolution layers \cite{shi16pixelshuffle}. 
The discriminator also consists of blocks of 2D convolutional layers followed by instance normalization and GLU activation.
\\
MaskCycleGAN-VC introduces a self-supervised auxiliary training task that aims to better capture the time-frequency structure during mel-spectrogram conversion. 
The auxiliary task consists of randomly masking frames of the input mel-spectrogram, which need to be filled by the model during training. 
We also adopted the masking task in our method. 
%\vspace{-2mm}
\subsection{Proposed method}\label{ssec:our_method}
%\vspace{-1mm}
\begin{figure}[t]
  \centering
  \includegraphics[width=1.0\linewidth]{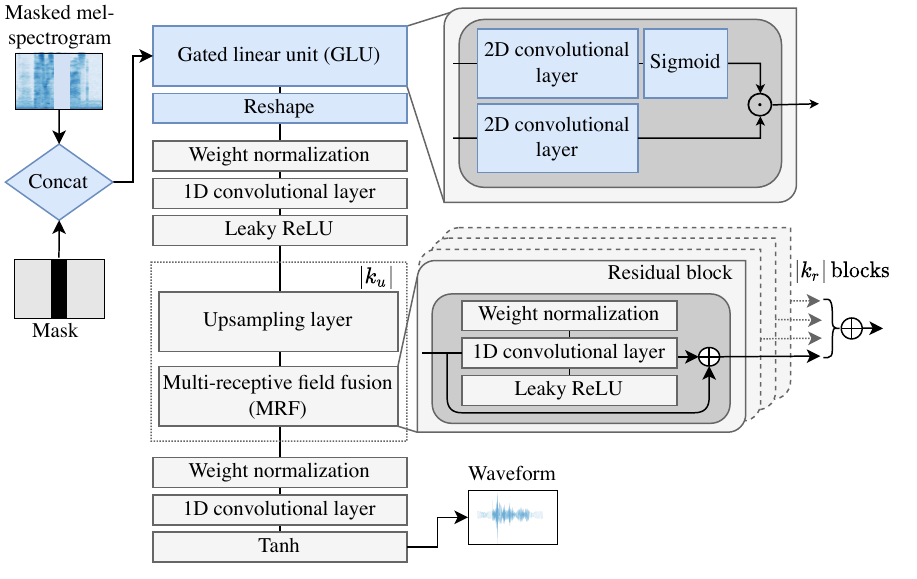}
  \vspace{-8mm}
  \caption{Generator of the proposed model. Blue-filled components have been added to the original HiFi-GAN generator. A GLU feature encoder block consisting of 2D convolutional layers operates on the concatenation of the masked mel-spectrogram and the mask itself. The resulting feature maps are reshaped and passed to a 1D convolutional layer before upsampling them in $|k_u|$ blocks.}
  \label{fig:generator}
  %\vspace{-4mm}
\end{figure}
\begin{figure}[h!]
  %\centering
  \includegraphics[width=0.95\linewidth]{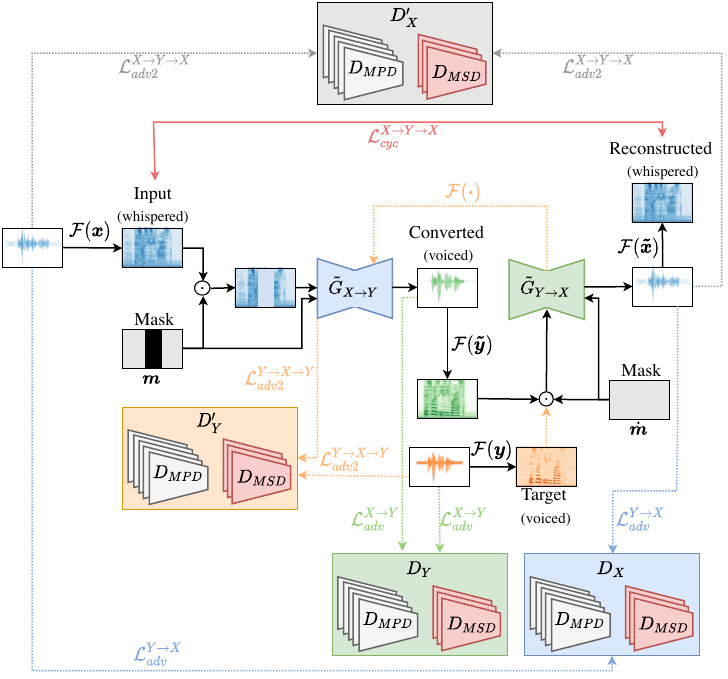}
  \vspace{-4mm}
  \caption{Overall audio generation and feedback process of our method. 
  Solid black lines represent feature transformations and flow of features through the two masked generators ($\tilde{G}_{X \rightarrow Y}$ and $\tilde{G}_{Y \rightarrow X}$). Dashed lines represent the flow of features to provide adversarial feedback. 
  Each of the four discriminators consists of two sub-discriminators ($D_{MPD}$ and $D_{MSD}$) receiving raw waveform inputs from varying sources. The solid red line indicates the features used to compute the forward cycle-consistency loss $\mathcal{L}_{cyc}^{X \rightarrow Y \rightarrow X}$. The losses $\mathcal{L}_{id}$ and $\mathcal{L}_{c y c}^{Y \rightarrow X \rightarrow Y}$ are excluded for clarity.}
  \label{fig:process}
%\vspace{-5mm}
\end{figure}

The goal is to learn a mapping $G_{X \rightarrow Y}$ between source waveforms $\mathbf{x} \in X$ and target waveforms $\mathbf{y} \in Y$ without parallel supervision. 
To achieve this goal, we train a CycleGAN \cite{zhu17cyclegan}, with an architecture based on HiFi-GAN \cite{kong20hifigan}. 
The generator receives randomly masked mel-spectrograms, as well as the mask itself, and uses a GLU block to encode the input features before upsampling them to the raw waveform. 
The components of the generator are illustrated in Figure \ref{fig:generator}. 
In addition to the standard adversarial loss \cite{goodfellow14gan} and the cycle-consistency loss \cite{zhu17cyclegan}, we use an additional identity-mapping loss \cite{taigman17identloss}, as well as a second adversarial loss \cite{kaneko19cycleganvc2}. 
The overall training objective is similar to \cite{kaneko19cycleganvc2,kaneko20cycleganvc3,kaneko21maskcyclevc}. 
The main difference is that the training objectives for the generator and the discriminator follow \cite{zhu17cyclegan,kong20hifigan}, which replace the binary cross-entropy terms of the original GAN objectives \cite{goodfellow14gan} with least squares loss functions \cite{mao17leastsquares} to stabilize the training procedure. 
The discriminators are trained using information from feature maps of the $k$ components in the multi-scale discriminator and the multi-period discriminator \cite{kong20hifigan}.
Furthermore, an additional transformation $\mathcal{F}$ is used to map from the raw audio waveform to the corresponding mel-spectrogram representation to allow loss computation in the frequency domain. 
The process of waveform generation and flow of adversarial feedback are illustrated in Figure \ref{fig:process}. 
The adversarial losses for the discriminator $\mathcal{L}_{adv}^{X \rightarrow Y}(D_{Y}; G_{X \rightarrow Y})$ and the generator $ \mathcal{L}_{adv}^{X \rightarrow Y}(G_{X \rightarrow Y} ; D_{Y})$ are defined as:
%%\vspace{-1mm}
\begin{equation}\label{eq:adv_loss_d1}
    \begin{aligned}
    &\mathcal{L}_{adv}^{X \rightarrow Y}(D_{Y}; G_{X \rightarrow Y})=\\
    &\mathbb{E}_{(\boldsymbol{s}, \boldsymbol{y})}\left[(D_{Y}(\boldsymbol{y})-1)^2+(D_{Y}(G_{X \rightarrow Y}(\boldsymbol{s})))^2\right]
\end{aligned}
%%\vspace{-2mm}
\end{equation}
\begin{equation}\label{eq:eq:adv_loss_g1}
    \mathcal{L}_{adv}^{X \rightarrow Y}(G_{X \rightarrow Y} ; D_{Y})=\mathbb{E}_{\boldsymbol{s}}\left[(D_{Y}(G_{X \rightarrow Y}(\boldsymbol{s}))-1)^2\right],
\end{equation}
where $\boldsymbol{s}= \mathcal{F}(\boldsymbol{x})$ denotes the mel-spectrogram of the input audio in domain $X$ and $\boldsymbol{y}$ denotes the target audio in domain $Y$. 
The backward discriminator $D_X$  (cf. \cite{zhu17cyclegan}) is trained with outputs from the backward generator $G_{Y \rightarrow X}$ using $\mathcal{L}_{adv}^{Y \rightarrow X}(D_{X}; G_{Y \rightarrow X})$. 
The backward generator $G_{Y \rightarrow X}$ is trained with outputs from the discriminator $D_X$ using $ \mathcal{L}_{adv}^{Y \rightarrow X}(G_{Y \rightarrow X}; D_{X})$. 

% $\mathcal{L}_{Adv}^{X \rightarrow Y}(D_{Y}; G_{X \rightarrow Y})$ and the generator $ \mathcal{L}_{Adv}^{X \rightarrow Y}(G_{X \rightarrow Y} ; D_{Y})$ 
The mappings between the source and target domains $X$ and $Y$ should be cycle-consistent, i.e., for each waveform $\boldsymbol{x}$ from domain $X$, the transformation cycle should be able to restore $\boldsymbol{x}$ back to its original form (and vice versa for $\boldsymbol{y} \in Y$): 
%%\vspace{-1mm}
\begin{equation}
\boldsymbol{x} \longmapsto G_{X \rightarrow Y}(\mathcal{F}(\boldsymbol{x})) \longmapsto G_{Y \rightarrow X}(G_{X \rightarrow Y}(\mathcal{F}(\boldsymbol{x}))) \approx \boldsymbol{x}.
%%\vspace{-1.0mm}
\end{equation}
The loss $\mathcal{L}_{c y c}^{X \rightarrow Y \rightarrow X}$ is used to implement the forward cycle-consistency constraint:
%\vspace{-1mm}
\begin{equation}\label{eq:cyc_loss}
    \mathcal{L}_{\text{cyc}}^{X \rightarrow Y \rightarrow X}
    =\mathbb{E}_{\boldsymbol{s}}\left[ || \mathcal{F} \left( G_{Y \rightarrow X}\left(G_{X \rightarrow Y}(\boldsymbol{s})\right) \right) -\boldsymbol{s}||_1\right]
    %\vspace{-1mm}
\end{equation}
Similarly, $\mathcal{L}_{c y c}^{Y \rightarrow X \rightarrow Y}$ is used for the backward mapping $G_{X \rightarrow Y}\left( G_{Y \rightarrow X}\left( \mathcal{F}(\boldsymbol{y}) \right) \right)$. 
\\
The identity loss term $\mathcal{L}_{i d}^{X \rightarrow Y}$ encourages $G_{X \rightarrow Y}$ to be the identity mapping for all $\boldsymbol{y} \in Y$:
%%\vspace{-2mm}
\begin{equation}\label{eq:ident_loss}
    \mathcal{L}_{i d}^{X \rightarrow Y}
    =\mathbb{E}_{\mathcal{F}(\boldsymbol{y})}\left[||\mathcal{F} \left( G_{X \rightarrow Y}(\mathcal{F}(\boldsymbol{y})) \right) -\mathcal{F}(\boldsymbol{y})||_1\right]
    %%\vspace{-1mm}
\end{equation}
For the backward generator $G_{Y \rightarrow X}$, the identity loss $\mathcal{L}_{i d}^{Y \rightarrow X}$ is used in the same way. 

A second adversarial loss $\mathcal{L}_{adv2}^{X \rightarrow Y \rightarrow X}$ operating on the features obtained after a full forward or backward cycle, and an additional discriminator $D^{\prime}_X$ are introduced to mitigate oversmoothing caused by the cycle-consistency constraints: 
%%\vspace{-2mm}
\begin{equation}\label{eq:adv_loss_d2}
    \begin{aligned}
    &\mathcal{L}_{adv2}^{X \rightarrow Y \rightarrow X}\left( D^{\prime}_{X}; G_{X \rightarrow Y}, G_{Y \rightarrow X}\right)=\\
    &\mathbb{E}_{(\boldsymbol{s}, \boldsymbol{x})}\left[(D^{\prime}_{X}(\boldsymbol{x})-1)^2
    +(D^{\prime}_{X}(G_{Y \rightarrow X}(G_{X \rightarrow Y}(\boldsymbol{s}))))^2\right]
\end{aligned}
%%\vspace{-1mm}
\end{equation}
\begin{equation}\label{eq:eq:adv_loss_g2}
\begin{aligned}
    &\mathcal{L}_{adv2}^{X \rightarrow Y \rightarrow X}(G_{Y \rightarrow X} ; G_{X \rightarrow Y} , D^{\prime}_{X})=\\
    &\mathbb{E}_{\boldsymbol{s}}\left[(D^{\prime}_{X}(G_{Y \rightarrow X}(G_{X \rightarrow Y}(\boldsymbol{s})))-1)^2\right],
\end{aligned}
%%\vspace{-1mm}
\end{equation}
where the discriminator $D^{\prime}_X$ distinguishes between the reconstructed \\ $G_{Y \rightarrow X}(G_{X \rightarrow Y}(\mathcal{F}(\boldsymbol{x})))$ and the real $\boldsymbol{x}$. 
Similarly, $\mathcal{L}_{adv2}^{Y \rightarrow X \rightarrow Y}(G_{X \rightarrow Y} ; G_{Y \rightarrow X} ; D^{\prime}_{Y})$ is used for the backward mapping with the additional discriminator $D^{\prime}_Y$. 

The discriminators are trained on the sum of the adversarial losses for each of the $K$ MPD/MSD discriminator blocks:
%%\vspace{-2mm}
\begin{equation}
\begin{aligned}
    &\mathcal{L}_{D} = \sum_{k=1}^K \left(
    \mathcal{L}_{adv}^{X \rightarrow Y}(D_{Y}^{(k)}; G_{X \rightarrow Y}) \right) \\
    &+ \sum_{k=1}^K \left( \mathcal{L}_{adv2}^{X \rightarrow Y \rightarrow X}(D_{X}^{\prime(k)}; G_{X \rightarrow Y}, G_{Y \rightarrow X}) \right).
\end{aligned}
%%\vspace{-2mm}
\end{equation}
The overall generator objective $\mathcal{L}_{G}$ is given by:
%%\vspace{-3mm}
\begin{equation}\label{eq:full_loss}
    \begin{aligned}
        &\mathcal{L}_{G}
        =\sum_{k=1}^K \left( \mathcal{L}_{a d v}^{X \rightarrow Y}(G_{X \rightarrow Y} ; D_{Y}^{(k)})
        +\mathcal{L}_{a d v}^{Y \rightarrow X}(G_{Y \rightarrow X} ; D_{X}^{(k)})\right)\\
        &+\lambda_{c y c}\left(\mathcal{L}_{c y c}^{X \rightarrow Y \rightarrow X}
        +\mathcal{L}_{c y c}^{Y \rightarrow X \rightarrow Y}\right)
        +\lambda_{i d}\left(\mathcal{L}_{i d}^{X \rightarrow Y}
        +\mathcal{L}_{i d}^{Y \rightarrow X}\right) \\
        &+\sum_{k=1}^K \left( \mathcal{L}_{a d v 2}^{X \rightarrow Y \rightarrow X}(G_{Y \rightarrow X} ; G_{X \rightarrow Y} , D_{X}^{\prime(k)}) \right) \\
        &+\sum_{k=1}^K \left( \mathcal{L}_{a d v 2}^{Y \rightarrow X \rightarrow Y}(G_{X \rightarrow Y} ; G_{Y \rightarrow X} , D_{Y}^{\prime(k)}) \right) ,
\end{aligned}
%%\vspace{-3mm}
\end{equation}
where $\lambda_{cyc}$ and $\lambda_{id}$ are scalars for weighting. 
%%\vspace{-3mm}
\subsubsection{Training with feature masking}
Kaneko et al. \cite{kaneko21maskcyclevc} proposed an additional unsupervised training task that helps to learn the harmonic structure of the mel-spectrogram features. 
The source mel-spectrogram $\boldsymbol{s} = \mathcal{F}(\boldsymbol{x})$, is augmented using a binary temporal mask $\boldsymbol{m} \in M$ of same size as $\boldsymbol{s}$. 
A randomly selected contiguous region consisting of $n$ frames is mapped to 0, while the remaining frames are mapped to 1. 
The mask is applied to each $\boldsymbol{s}$ using element-wise multiplication: $\boldsymbol{\tilde{s}} = \boldsymbol{s} \cdot \boldsymbol{m}$. 
\\
This procedure effectively creates missing frames on the input features, which need to be filled by the generator. 
The mask is also used as an additional input to the generator to provide information about the frames that need to be filled. 
It is passed to the forward generator and subsequently concatenated with the mel-spectrogram on channel-level. 
The forward cycle generates $\boldsymbol{\tilde{y}} = \tilde{G}_{X \rightarrow Y}(\boldsymbol{\tilde{s}}, \boldsymbol{m})$. 
The backward transformation $\tilde{G}_{Y \rightarrow X}$ employs an all-ones matrix $\dot{\boldsymbol{m}}$ to reconstruct $\boldsymbol{\tilde{x}}$:
%%\vspace{-1mm}
\begin{equation}
    \boldsymbol{\tilde{x}} = \tilde{G}_{Y \rightarrow X}\left( \mathcal{F}\left( \boldsymbol{\tilde{y}} \right), \dot{\boldsymbol{m}} \right)
    %%\vspace{-1mm}
\end{equation}
%%\vspace{-1mm}
To optimize the cycle-consistency loss, the generator $\tilde{G}_{X \rightarrow Y}$ is encouraged to learn an estimate of the missing frames by using information about the neighboring frames in a self-supervised manner. 
We found that the frame-filling task can greatly improve the time-frequency structure of the converted whispered inputs. 
Recovering the correct harmonic structure from whispered speech is especially important, since it is almost completely absent in most samples. 
%\vspace{-3mm}
\section{Experiments}
%\vspace{-2mm}
We conducted seven experiments on the whispered TIMIT (wTIMIT) corpus \cite{lim10wtimit}. The first three experiments were designed to allow a comparison of our proposed method with the following systems: (1) MaskCycleGAN-VC \cite{kaneko21maskcyclevc}, (2) original HiFi-GAN \cite{kong20hifigan} trained with DTW-aligned parallel input features, (3) NVC-Net \cite{bac22nvcnet}. 
The experiments (4) to (7) are different variants of the proposed model to gauge the importance of the individual key components: (4) regular cycle-consistent training only (i.e., no masking, no additional $\mathcal{L}_{adv2}$, and no GLU feature encoder), (5) same as (4) but with masking added to the training procedure, (6) masking and additional $\mathcal{L}_{adv2}$ used during training, (7) full model with masking, additional $\mathcal{L}_{adv2}$ and GLU feature encoder. 
\\
Experiment (2) was inspired by \cite{wagner22whisper}.
The HiFi-GAN was trained using the hyperparameters and loss functions from \cite{kong20hifigan} on parallel utterance pairs, which were time-aligned in a preprocessing step. 
NVC-Net (exp. (3)) was trained using the default hyperparameters from \cite{bac22nvcnet}. 
\\
Except for (2), all experiments were conducted in a non-parallel setting, i.e., the source and target utterances were not required to have the same linguistic content. 
The utterances and input frames from these utterances were selected randomly in each training step. 
Additionally, we used the experimental setups (1) and (7) to compare the conversion quality of MaskCycleGAN-VC and our method on the VCTK dataset \cite{vctk19}.  
%\vspace{-2mm}
\subsection{Data}
%\vspace{-1mm}
The wTIMIT corpus uses approximately 450 phonetically compact sentences from the TIMIT \cite{garolfo93timit} database. 
Each speaker utters the same sentences in a regular (voiced) and in a whispered manner. 
The corpus consists of 24 female speakers and 25 male speakers from two main accent groups (Singaporean-English and North-American English).
We used 10 speakers (5 female and 5 male) from both accent groups for training, i.e., we trained 10 individual speaker-dependent models in each experiment.  
We excluded 20 randomly selected utterances from each speaker (200 utterances in total) as a test set. 
The test set is consistent across all 10 speakers, i.e., the spoken content in those 20 utterances is the same for each of the 10 selected speakers. 
The audio format is 16-bit PCM with a sample rate of 44.1 kHz. 
We applied volume normalization and trimming of leading and trailing silences to each audio sample to reduce the impact of noise artifacts.\\ 
The VCTK corpus \cite{vctk19} contains 16-bit FLAC recordings sampled at 48 kHz from 109 English speakers, who read about 400 sentences each.  
We randomly selected 6 speakers (3 male, 3 female) and also excluded 20 utterances from each speaker as a test set.\\ 
We reduced the sample rate of both corpora to 22.05 kHz and extracted 80-dimensional mel-spectrograms with a window length of 1024 and hop length of 256 samples. 
%\vspace{-2mm}
\subsection{Training setup}
%%\vspace{-2mm}
\subsubsection{MaskCycleGAN-VC}
%\vspace{-1mm}
We adopted most training hyperparameters from \cite{kaneko21maskcyclevc}.\\ 
Mel-spectrograms were normalized using the training set statistics. 
The models were trained for 50k iterations (batch size 8) using the Adam
optimizer \cite{kingma15adam} with learning rates of $2\times10^{-4}$ for the generator and $10^{-4}$ for the discriminator. 
The momentum terms $\beta_1$ and $\beta_2$ were set to $0.5$ and $0.99$, respectively. 
The input length was set to 64 frames with a maximum of 25 consecutive frames being masked. 
We used $\lambda_{cyc} = 10$ and $\lambda_{id}=5$. 
The identity loss was only used for the first 1k iterations. 
To ensure a fair comparison with our method, we synthesized the mel-spectrograms converted with MaskCycleGAN-VC using a HiFi-GAN vocoder pretrained on the VCTK dataset \cite{vctk19}, instead of the MelGAN \cite{kundan19melgan} vocoder suggested in \cite{kaneko21maskcyclevc}. 
%%\vspace{-1mm}
\subsubsection{Our method}
%\vspace{-2mm}
We employed 3 MSD and 5 MPD blocks in each discriminator. 
The GLU feature encoder yielded 64 channels and used kernel sizes $(5,15)$. 
Upsampling was achieved using 4 blocks of transposed 1D convolutional layers followed by MRF modules. 
The upsampling rates, i.e, the stride values were set to $u \in \lbrace 8, 8, 2, 2 \rbrace$. 
The kernel sizes $k_u$ of the transposed convolutions were set to $k_u = 2 \times u$. 
We used 3 residual blocks in each MRF module with kernel sizes $k_r \in \lbrace 2,7,11 \rbrace$ and dilation rates $d_r \in \lbrace 1,3,5 \rbrace$. 

The models were trained for 50k iterations (batch size 8) using the Adam optimizer with $\beta_1 = 0.5$ and  $\beta_2 = 0.99$. 
An exponential learning rate scheduler with a rate decay factor of $0.999$ in every epoch and an initial learning rate of $2 \times 10^{-4}$ was used for the generators and discriminators alike. 
We used $\lambda_{cyc} = 10$ and $\lambda_{id}=5$ as weights for the cycle-consistency loss and the identity loss, respectively. 
The input length was also 64 frames with a maximum of 25 consecutive frames being masked. 
%\vspace{-4mm}
\subsection{Objective evaluation}
%\vspace{-2mm}
We employed three commonly used measures to assess the conversion results: 
mel-cepstral distortion (MCD), frequency-weighted segmental signal-to-noise ratio (fwSNRseg), and root mean squared error (RMSE) of $log \, F0$. 
To obtain MCD and RMSE, we extracted 34 Mel-cepstral coefficients (MCEPs) at a frame period of 5 ms for each recording sampled at 22.05 kHz using the WORLD analysis system \cite{masanori16world}. 
Similar to \cite{parmar19whisper,wagner22whisper,Wagner2024}, we computed the RMSE of $log \, F0$ values between the voiced and converted speech signals, after aligning all frames in the utterance via DTW. 
Only the voiced regions of the target signal were considered in the computation. 
The results are summarized in Table \ref{tab:objective_eval}. 
All experiments yielded improvements over the whispered input condition. 
Using the original HiFi-GAN on parallel input data (exp. (2)) led to the best results in terms of MCD and fwSNRseg, but the reconstruction quality of $log \, F0$ was worse than most other experiments. 
The lowest overall RMSE was achieved with our full model. 
%\vspace{-4mm}
\begin{table}[!htb]
\setlength{\tabcolsep}{12pt} % Default value: 6pt
\centering
\caption{\small{Results for objective quality measures on the wTIMIT test set. 
Arrows indicate whether lower ($\downarrow$) or higher ($\uparrow$) values are better for each measure.}}
\label{tab:objective_eval}
%\vspace{-2mm}
\scalebox{0.99}{
\begin{tabular}{llccc}
\hline 
\textbf{\#} &
\textbf{Experiment} & 
\makecell{\textbf{MCD} \\ {[}dB{]}$\downarrow$} &
\makecell{\textbf{fwSNRseg} \\ {[}dB{]}$\uparrow$} &
\makecell{\textbf{RMSE} \\ {[}log F0{]}$\downarrow$} \\
\hline
1 & MaskCycleGAN-VC    & 8.588 &  0.956 & 18.888 \\
2 & HiFi-GAN + DTW     & \textbf{7.891} & \textbf{2.531} &  19.927 \\
3 & NVC-Net & 8.683 & 1.074 & 19.577 \\
4 & Ours \scriptsize{(cycle-consistent only)}  & 8.175 & 1.602 &  19.971 \\
5 & \hspace{3mm}+ feature masking & 8.067 & 2.046 & 20.342\\
6 & \hspace{3mm}+ $\mathcal{L}_{adv2}$ & 8.183 & 1.619 &  18.996  \\
7 & \hspace{3mm}+ GLU feature encoder & 8.028 & 2.056  &  \textbf{18.257} \\
\hline
 {} & Whispered           & 9.523 & -0.768 & 27.991 \\
\hline
\end{tabular}
}
%%\vspace{-5mm}
\end{table}
% 1.074361	19.577230	8.683020
% cycle_mask_2d_config_v2_iter_80k	2.059872	18.341974	8.028249
%%\vspace{-4mm}
\subsection{MOS prediction}\label{ssec:mospred}
%%\vspace{-1mm}
% NOTE: VCC 2016 evaluates in terms of target speaker similarity and
% naturalness (200 listeners in a controlled environment)
We utilized a pre-trained model for automatic prediction of mean opinion scores (MOS) \cite{tseng21mospred}. 
%The model relies on wav2vec 2.0 \cite{baevski20w2v2} as a feature extractor and uses a stack consisting of attention pooling \cite{safari20attpooling}, a bias network similar to MBNET \cite{leng21mbnet}, and range clipping to predict mean opinion scores from waveform data.
%We used a model pre-trained on Voice Conversion Challenge 2018 (VCC 2018) \cite{lorenzo18vcc18} and Voice Conversion Challenge 2016 (VCC 2016) \cite{toda16vcc16} data in our experiments. 
The results on the wTIMIT corpus in Table \ref{tab:mos_pred} show that our full model (exp. (7)) achieved a score closest to the voiced target speech.  
Removing the GLU feature encoder (exp. (6)) still achieves results comparable to to the baseline. 
\begin{table}[!htb]
%\vspace{-2mm}
\setlength{\tabcolsep}{18pt} % Default value: 6pt
\centering
\caption{\small{Average MOS prediction, standard deviation, and 95\% confidence interval on the wTIMIT test set.}}
% i.e., 200 utterances (20 utterances per speaker).
\label{tab:mos_pred}
%\vspace{-2mm}
\scalebox{0.99}{
\begin{tabular}{llcc}
\hline
\textbf{\#} &
\textbf{Experiment} & 
\textbf{MOS pred.} & 
\textbf{95\% CI}
%\textbf{\# params} 
\\
\hline
1 & MaskCycleGAN-VC  & $3.07 \pm 0.11$    & $0.08$ \\
2 & HiFi-GAN + DTW   & $2.91 \pm 0.12$    & $0.09$ \\
3 & NVC-Net          & $2.79 \pm 0.08$    & $0.05$ \\
4 & Ours \scriptsize{(cycle-consistent only)}  & $2.73 \pm 0.08$    & $0.06$ \\
5 & \hspace{3mm}+ feature masking & $2.94 \pm 0.11$ &  $0.08$ \\
6 & \hspace{3mm}+ $\mathcal{L}_{adv2}$ & $3.00 \pm 0.09$   & $0.06$\\
7 & \hspace{3mm}+ GLU feature encoder & $\boldsymbol{3.16}  \pm 0.11$ & $0.08$\\
\hline
{} & Voiced           & $3.47  \pm 0.09$  & $0.07$ \\
{} & Whispered         & $2.68  \pm 0.10$    & $0.07$ \\
\hline
\end{tabular}
}
\end{table}
We performed conventional VC experiments on the VCTK dataset, to further analyze the effectiveness of our method. 
Table \ref{tab:mos_pred_vctk} compares our full model (exp. (7)) to MaskCycleGAN-VC. 
Our method achieved slightly better MOS values than MaskCycleGAN-VC in all four experiments and yielded the highest overall MOS of 3.40 in the female-to-male ($F \mapsto M$) conversion experiment. 
%%\vspace{-3mm}
\begin{table}[!htb]
\setlength{\tabcolsep}{12pt} % Default value: 6pt
\centering
\caption{\small{MOS prediction results on VCTK test data.}}
\label{tab:mos_pred_vctk}
%\vspace{-2mm}
\scalebox{0.99}{
\begin{tabular}{cccc}
\hline
%{} & \multicolumn{3}{c}{\textbf{MOS prediction}} \\
\textbf{Exp.} & 
% \multicolumn{2}{c}{}
\textbf{MaskCycleGAN-VC} & 
\makecell{\textbf{Ours} \\ (\#7)} &
\makecell{\textbf{Ground truth} \\ (target)}
\\
\hline
$F \mapsto F$& $3.30 \pm 0.07$  & $3.38 \pm 0.10$  & $3.53 \pm 0.06$ \\
$M \mapsto M$& $3.23 \pm 0.09$  & $3.30 \pm 0.16$  & $3.45 \pm 0.07$  \\
$F \mapsto M$& $3.32 \pm 0.09$  & $3.40 \pm 0.12$  & $3.51 \pm 0.06$  \\
$M \mapsto F$& $3.36\pm 0.08$ & $3.38 \pm 0.15$   & $3.52 \pm 0.07$  \\
\hline
\end{tabular}
}
\vspace{-2mm}
\end{table}
%%\vspace{-1mm}
\subsection{Subjective evaluation}
\begin{figure}[h!]
  \centering
  \includegraphics[width=0.8\linewidth]{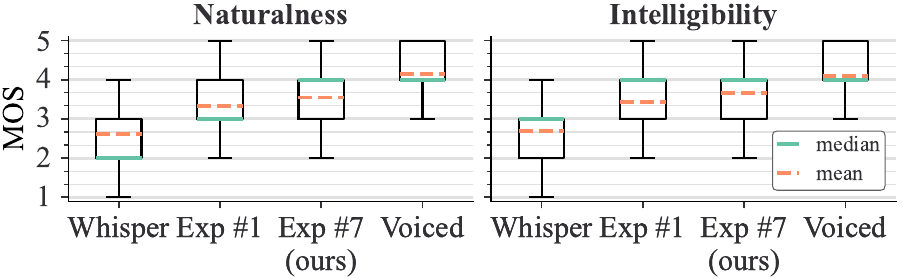}
  \vspace{-3mm}
  \caption{Subjective human evaluation results (MOS) for naturalness and intelligibility.}
  \label{fig:subj_eval}
  \vspace{-3mm}
\end{figure}

%%\vspace{-1mm}
We conducted two separate listening tests on \textit{naturalness} and \textit{intelligibility}, asking human listeners to choose on a 5-point scale with choices from ``excellent'' to ``bad'', to further assess the performance of our method. 
We used 5 utterances from each of the 10 speakers in the wTIMIT test set (50 utterances in total).  
We also asked the participants to rate the underlying whispered and voiced speech samples. 
We ensured that at least 30 participants completed the listening test. 
The results are illustrated in Figure \ref{fig:subj_eval}. 
Similar to the MOS prediction results in Section \ref{ssec:mospred}, our method outperformed MaskCycleGAN-VC in terms of both naturalness ($\mu = 3.54$ vs. $\mu = 3.32$) and intelligibility ($\mu = 3.66$ vs. $\mu = 3.43$). 
Both systems yielded higher scores than the whispered speech ($\mu = 2.62$ naturalness and $\mu = 2.69$ intelligibility), but lower scores than the voiced target speech ($\mu = 4.15$ naturalness and $\mu = 4.09$ intelligibility).
\vspace{-5mm}
\section{Conclusions}
%%\vspace{-2mm}
We describe an efficient method to convert whispered speech features directly into voiced audio waveforms. 
Our method utilizes components from MaskCycleGAN-VC and HiFi-GAN to recover the harmonic structure and to generate the waveform in a single model, while avoiding artifacts caused by imperfect time alignments. 
Experimental results indicate that our unified approach maintains competitive results in both whispered speech conversion and conventional VC. 

\bibliographystyle{splncs04}
\footnotesize{
\bibliography{refs}
}
\end{document}